\documentclass[sigconf]{acmart}

\copyrightyear{2026}
\acmYear{2026}
\setcopyright{cc}
\setcctype{by}
\acmConference[AVI '26]{Proceedings of the 2026 International Conference on Advanced Visual Interfaces}{June 08--12, 2026}{Venice, Italy}
\acmBooktitle{Proceedings of the 2026 International Conference on Advanced Visual Interfaces (AVI '26), June 08--12, 2026, Venice, Italy}
\acmDOI{10.1145/3811427.3811443}
\acmISBN{979-8-4007-2342-1/2026/06}

\graphicspath{{figures/}{pictures/}{images/}{./}}

\usepackage{enumitem}

\usepackage[fixed]{fontawesome5}

\usepackage{ellipsis}

\PassOptionsToPackage{hyphens}{url}

\usepackage{tabularx}
\usepackage{multirow}

\usepackage{xspace}

\usepackage{subcaption}

\usepackage{soul}

\usepackage{csquotes}

\definecolor{new-green}{rgb}{0.05,0.50,0.05}
\newcommand{\del}[1]{} 

\begin{document}


\title[TombWriter]{TombWriter: Scaffolding Story Archeology through Beat-Level Interaction in Human-AI Co-Writing}

\author{Hugo Andersson}
\email{hugo@cs.au.dk}
\orcid{0009-0009-1347-2836}
\affiliation{
  \institution{Aarhus University}
  \city{Aarhus}
  \country{Denmark}
}

\author{Niklas Elmqvist}
\email{elm@cs.au.dk}
\orcid{0000-0001-5805-5301}
\affiliation{
  \institution{Aarhus University}
  \city{Aarhus}
  \country{Denmark}
}

\renewcommand{\shortauthors}{Andersson \& Elmqvist}

\begin{abstract}
    The dominant paradigm for LLM interaction in AI co-writing uses disposable prompts that vanish after use.
    This may lead to imprecise results, cumbersome workflows, and diminished author agency and ownership.
    We propose \textit{LLM-based story archeology}, where prompts serve as a hierarchical story instrument refined over time to extract the writer's intended story.
    Drawing on the fossil theory of storytelling, where stories exist as latent structures that writers excavate through their craft, this approach supports agency and ownership through high involvement and control.
    Writers work at the level of story beats rather than prose.
    They generate character actions in scenes to discover emergent possibilities, simulated by the LLM or directly nudged, then edit resulting beats to refine scenes iteratively.
    Prose is generated from beats based on style and genre, separating structure from style.
    We developed \textsc{TombWriter}, a web-based tool that visualizes stories as navigable cards---characters, scenes, and beats---through a five-stage narrative pipeline.
    We conducted a qualitative study with five experienced writers who used the system over three days.
    Through semi-structured interviews, we found that writers framed AI as a generation engine rather than collaborator, claimed ownership while reporting voice loss, and valued the system for structural discovery rather than prose production.
    We contribute the story archeology approach, the TombWriter system, and qualitative findings on beat-level human-AI co-writing.
\end{abstract} 



\maketitle


\section{Introduction}

Generative AI in general, and Large Language Models (LLMs) in particular, have become a great enabling technology for democratizing creative writing~\cite{Lee2024designspace}.
However, the dominant LLM chatbot interfaces today are fundamentally prompt-based~\cite{DBLP:conf/nips/BrownMRSKDNSSAA20}, supporting the co-writing process as a sequence of disposable queries and responses that yield LLM-generated prose.
The end result is that human writers often perceive a lack of agency in the creative workflow, and accordingly feel that their sense of ownership of the finished product is diminished or even lost~\cite{hoque2023hallmark, Lee2024designspace, Reza2025cowriting, draxler_ai_2024}. 
When a throwaway prompt can add hundreds of badly integrated words to your story, is that story even yours~\cite{Kreminski2024dearth}?

In this paper, we propose \textit{LLM-based story archeology} as an approach to ensure writer agency in AI creative co-writing that aims to address these concerns.
We use the term ``story archeology'' in deference to writer Stephen King's ``fossil theory'' of storytelling~\cite{king2000onwriting} where stories are not invented \textit{ex nihilo}, but already exist in some latent form, like fossils buried underground. 
The role of the writer is, accordingly, similar to an archeologist at a dig site: to discover and excavate the story using available instruments.
Rather than focusing on generating prose, LLM-based story archeology enables writers to iteratively develop a \textit{hierarchical story instrument} that captures their intent as a persistent specification.

This instrument consists of characters and scenes, where each scene is a sequence of story beats that capture dramatic actions as abstract descriptions.
Scenes are generated beat by beat, either through LLM simulation of character actions—what writers call \textit{organic storytelling} or ``seat of your pants writing''—or through writer nudging.
The writer can manually edit each beat to their liking.
Because the creative process happens at the instrument level rather than prose, the actual prose can be generated later based on writing style, genre, and desired length.

To explore this approach, we developed \textsc{TombWriter}, a web-based tool that implements story archeology through a card-based interface for managing characters, scenes, and beats.
We conducted a qualitative study with five experienced writers who used TombWriter over three days to engage in a multi-session, intensive deployment, allowing them to develop their own stories.
Through semi-structured interviews, we examine how writers engage with beat-level authoring and surface insights around perceived control, the experience of character simulation, and the separation of structure from prose.

We contribute the following in this paper:
(1) LLM-based story archeology, an approach to AI co-writing centered on beat-level authoring and persistent story instruments;
(2) TombWriter, a web-based tool for exploring this approach; and
(3) qualitative findings from experienced writers on beat-level human-AI co-writing.

\section{Background}

Stories are central to the human condition, and accordingly, storytelling is a fundamentally human activity.
People write for self-expression, therapy, education, communication, and play.
However, few people ever become professional published authors.
Nevertheless, many people have ideas for stories, maybe even great ones, but lack the time, training, physical, or cognitive ability to follow through on them.
They may also lack the literary ability or language skills to write in a compelling fashion despite having the inclination and need.
But in the end, everyone's voice deserves to be heard, at least for their own self-expression and personal audience.

Here we give the necessary background on stories and storytelling. 
We discuss the anatomy of a story and how they are written by professional and amateur authors alike.
Finally, we discuss the new breed of AI co-writing tools that have emerged over the last few years to leverage large language models in service of the creative storytelling process.

\subsection{Plotting vs.\ Pantsing}

There are two fundamentally different approaches to creative storytelling: \textit{plotting} and \textit{pantsing}.
Plotting refers to authors who plan out the entire story in advance by writing detailed outlines.
Pantsing (``seat of your pants'' writing, or \textit{organic storytelling}) involves placing characters in conflict and determining the outcome by acting out character actions based on their goals, motivations, and personalities.
Plotting plans the outcome in advance; pantsing turns the creative process into a journey of discovery where the outcome is not initially known.
Hybrid approaches exist between these extremes~\cite{bell2011plot}.

Pantsing can be summarized by the E.L. Doctorow quote, \textit{``Writing is like driving at night in the fog. You can only see as far as your headlights, but you can make the whole trip that way.''}
Stephen King, a famous adherent to this approach, talks in his memoir \textit{On Writing}~\cite{king2000onwriting} of his distrust of plotting, largely because our real lives are plotless, and because relying on strict plots is antithetical to spontaneous creativity.
This view aligns with Suchman's theory of situated action~\cite{Suchman1987}, which argues that experts use plans as resources for action rather than predefined scripts, acting largely through improvisation and response to unfolding context.
King describes his process as creating rich, multi-dimensional characters with their own goals, personalities, and attributes, and then placing them in difficult situations.
The writing process becomes one of discovery as the characters act in a manner true to themselves, yielding the emerging story.
This character-driven approach and persistent structure supporting improvised action, inspires our beat-level story archeology method.

\subsection{LLMs for Creative Writing}

Before LLMs enabled generation, computational tools supported creative writing through analysis; identifying social biases in character portrayal~\cite{DBLP:conf/ACMdis/HoqueGE22} or analyzing fictional characters through NLP and visualization~\cite{DBLP:conf/ACMdis/HoqueGKE23}.
Large language models have fundamentally transformed computational support for creative writing~\cite{Lee2024designspace}.
Unlike the earlier tools focused on editing and revision, modern LLM-based systems can generate substantial original prose directly from prompts, enabling new forms of human-AI collaboration~\cite{Reza2025cowriting}.
Luther et al.\ found that users collaborating with ChatGPT showed high satisfaction and cognitive trust, primarily using prompts for content and information~\cite{Luther2024teaming}.
However, this generative power introduces critical challenges around writer agency, ownership, and creative control~\cite{hoque2023hallmark, Lee2024designspace, Reza2025cowriting, draxler_ai_2024}.
When AI generates hundreds of words in response to a simple prompt, questions arise about authorship and the writer's role in the creative process~\cite{Kreminski2024dearth}.

The most common approach employs turn-taking interfaces where writers alternate with the AI in generating prose.
Yuan et al.\ studied professional writers using Wordcraft, finding they valued AI for brainstorming but struggled with preserving authorial voice~\cite{DBLP:conf/iui/YuanCRI22}.
Similarly, Grigis and De Angeli's 7-month longitudinal study with professional playwrights found that despite using AI throughout, writers maintained that ``the creative effort was uniquely their own''---yet most ultimately rejected the technology as unsuitable for actual playwriting~\cite{grigis2024playwriting}.
TaleBrush uses sketching to control story generation~\cite{DBLP:conf/chi/ChungKYLAC22}, while ABScribe supports exploring multiple writing variations~\cite{DBLP:conf/chi/RezaLMDYMGLKW24}.
These systems focus on immediate text output rather than structural planning.

An alternative approach separates story structure from prose generation, addressing coherence issues through hierarchical planning.
Dramatron demonstrated this with prompt chaining. From log line to characters to plot outline to dialogue, they enable coherent scripts despite limited context windows~\cite{mirowski2023co}.
Yet a study with 15 theatre professionals revealed a persistent problem: writers valued Dramatron for inspiration but felt low ownership, seeing outputs as ``provocations'' rather than finished work.
Subsequent systems have pursued similar structural approaches: Re3 and DOC use recursive reprompting and outline control for longer stories~\cite{DBLP:conf/emnlp/YangTPK22, DBLP:conf/acl/YangKPT23}; WHAT-IF generates branching narratives through meta-prompting~\cite{huang2024if}; WhatELSE enables configurable abstraction levels with increased perceived control~\cite{lu2025whatelse}; and CALYPSO assists Dungeon Masters with contextual scene generation~\cite{DBLP:conf/aiide/ZhuMHC23}.
Visual approaches offer another avenue. Masson et al.\ found that writers using visual representations (entity graphs, timelines, location maps) could explore narrative structure more intuitively and specify edits by manipulating visual elements rather than prompting~\cite{masson2025visual}.
Crucially, they found that mental model alignment matters: writers whose natural thinking aligned with the visual representation benefited most, while others experienced increased cognitive load.
Yet writers still expressed reservations about AI overwriting their voice, preferring tools that highlight inconsistencies rather than rewrite text.
Across these systems, a pattern emerges: hierarchical generation and visual exploration improve structure and control, but do not resolve the ownership problem without closer writer involvement in the creative loop itself.

\subsection{Positioning TombWriter}

A central challenge in AI-assisted creative writing is not capability, but ownership. Writers using generative tools often produce satisfying text yet feel estranged from it~\cite{hoque2023hallmark, Lee2024designspace, Reza2025cowriting, draxler_ai_2024}.
This tension manifests in what Draxler et al.\ call the ``AI Ghostwriter Effect'': users do not perceive ownership of AI-generated text, yet hesitate to publicly acknowledge AI involvement~\cite{draxler_ai_2024}.
Chen and Chan's experimental work suggests collaboration modality itself shapes outcomes: when LLMs generate content for humans to refine (ghostwriter mode), users anchor to initial outputs and produce more homogeneous results---an effect particularly detrimental to experts~\cite{chen2024large}.
The disconnect intensifies when writers prioritize emotional investment over productivity: they want control precisely when turning ideas into words matters most~\cite{DBLP:conf/ACMdis/BiermannMY22}.
Crucially, this sense of ownership concentrates in the planning stage, where writers shape what the story \textit{is} before deciding how it reads~\cite{Reza2025cowriting}, precisely where beat-level authoring operates.

This body of work establishes that agency and ownership depend critically on when and how AI intervenes.
Existing LLM systems either generate prose directly (limiting structural control) or create plans in one shot (limiting iterative refinement).
TombWriter addresses this gap by combining character-driven simulation with persistent, refinable story structures at the beat level---keeping the human writer in the creative loop throughout.

\section{LLM-based Story Archeology}
\label{sec:framework}

We propose \textit{LLM-based story archeology} as a method for AI-assisted creative writing that addresses the agency and ownership problems inherent in prompt-based co-writing. 
The term draws on Stephen King's fossil theory of storytelling~\cite{king2000onwriting}, in which stories exist as latent structures that writers excavate rather than fabricate. 
In prompt-based systems, writers issue disposable queries that generate prose directly, creating an opaque transformation from intent to text that leaves writers feeling like curators rather than authors.
Story archeology inverts this paradigm: writers build a persistent \textit{story instrument}---a structured specification of characters, scenes, and story beats---that captures their creative intent explicitly. 
The LLM serves not as a prose generator but as a simulation engine that explores how characters would act given their established personalities and the current situation.
Writers direct this simulation through acceptance, rejection, nudging, and manual editing at the granularity of individual story beats. The actual prose becomes a secondary rendering step, generated from the beat sequences only when needed, allowing the same story structure to be expressed in different styles, genres, or voices. Story archeology rests on three core principles:

\paragraph{Persistence.}
Stories are created by iteratively refining a \textit{story instrument}: a hierarchical specification of the writer's storytelling intent.
Unlike disposable prompts, the instrument is persistent. Every character, scene, and story beat remains part of the specification and can be referenced, revised, and extended.
This persistence creates an accumulating context that the LLM uses to maintain coherence across scenes.
Character changes propagate forward, ensuring that a character who becomes distrustful in scene three remains distrustful in scene five unless something changes them again.
Furthermore, unlike one-off prompts that are quickly forgotten, the persistent instrument fosters a sense of ownership.

\paragraph{Abstraction.}
Writers work at the level of dramatic action, not prose.
A beat like ``Alice confronts Bob about the missing documents; Bob deflects by questioning Alice's motives'' captures the dramatic structure without committing to sentence-level choices.
This abstraction lets writers focus on story logic of what happens and why, separately from expression of how to phrase it engagingly.
The same beat sequence can generate literary fiction, pulp thriller prose, or a Hollywood screenplay, depending on the premise and characters.

\paragraph{Simulation.}
Rather than prompting the LLM to generate prose, writers use the LLM to simulate what characters would do given their established personalities and the current situation.
This organic storytelling approach treats characters as agents whose actions emerge from their attributes.
Writers can accept the simulated action, reject it and simulate again, or nudge the simulation toward a desired outcome.
The LLM becomes a tool for exploring story possibilities rather than a prose generator.
Because simulation shows writers why characters act as they do (based on visible attributes), and because writers retain multiple intervention points at the level of individual beats, they maintain agency over the creative decisions.

\begin{figure*}[t]
  \centering
  \includegraphics[width=\textwidth]{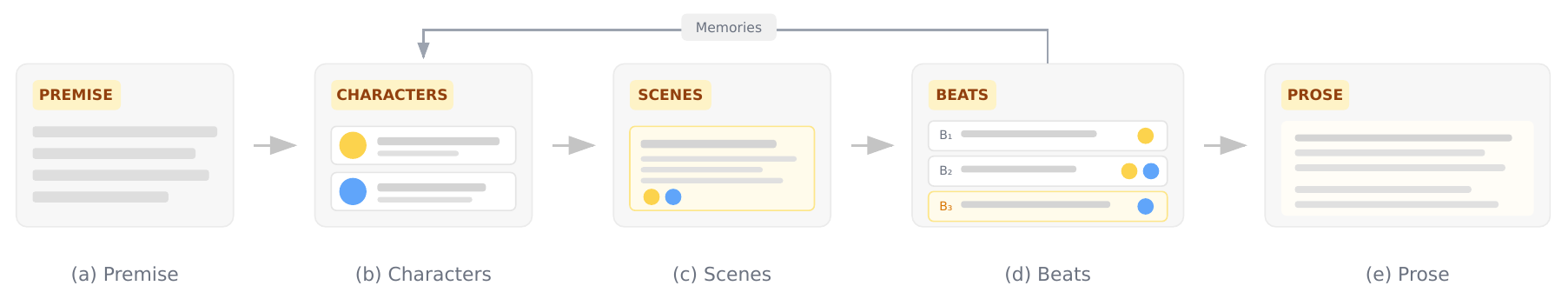}
  \Description{Horizontal five-stage pipeline diagram. Five colored rectangles are connected by arrows flowing left to right, labeled: Premise, Characters, Scenes, Beats, and Prose. The Premise card contains placeholder text lines. The Characters card shows two circular icons. The Scenes card shows a single rectangular element. The Beats card shows three horizontal beat rows, each with colored status indicators. The Prose card contains text lines. A curved arrow labeled "Memories" feeds back from the Beats stage to the Characters stage, indicating that characters accumulate memories from beats they participate in.}
  \caption{\textnormal{The TombWriter five-stage pipeline (schematic representation).} \textbf{(a)~Premise}\textnormal{: the writer articulates a high-level concept capturing the story's core conflict, theme, or dramatic question.} \textbf{(b)~Characters}\textnormal{: the writer creates character cards specifying traits, goals, and personality.} \textbf{(c)~Scenes}\textnormal{: the writer defines an initial situation and selects participating characters.} \textbf{(d)~Beats}\textnormal{: dramatic actions are generated iteratively through simulation, nudging, or manual editing. Characters accumulate memories from beats they participate in, enabling them to reference past events in subsequent scenes.} \textbf{(e)~Prose}\textnormal{: the final text is rendered from the beat sequence.}}
  \label{fig:pipeline}
\end{figure*}
  
\section{TombWriter}

To explore the story archeology approach, we developed TombWriter, a web-based prototype that enables writers to construct narratives through discrete story beats.
The design emphasizes two principles from story archeology: persistence (every creative decision remains visible and editable) and abstraction (writers work at the level of dramatic action, not prose).

TombWriter uses a card-based interface where each element of the story instrument---premise, characters, scenes, and beats---is represented as a manipulable card.
We chose cards as the primary metaphor because they make the story's building blocks tangible: writers can see their entire narrative structure at a glance, rearrange elements spatially, and maintain awareness of how individual beats connect to the broader story.
This contrasts with chat-based interfaces where prompts and outputs scroll past and disappear.

A typical workflow proceeds through five stages (Figure~\ref{fig:pipeline}).
Writers begin by articulating a premise, a high-level concept that captures the story's core conflict, theme, or dramatic question.
This premise anchors subsequent decisions and provides context for LLM generation.
Writers then create character cards, defining traits, goals, and personalities that will drive the characters' behavior during the simulation.
Next, they create scene cards, selecting participating characters and describing an initial situation.
Within each scene, writers generate beats iteratively, either through LLM simulation, nudging toward desired outcomes, or manual authoring. 
Finally, once the beat structure captures the intended story, prose can be rendered in the writer's chosen style and genre. 

\subsection{The Story Instrument}

The story instrument in TombWriter consists of a structured hierarchy. Writers first create \textit{characters} with descriptions, personality traits (rated on scales), and goals expressed in natural language. Characters accumulate memories from beats they participate in, enabling them to reference past events in subsequent scenes. Writers then create \textit{scenes} by specifying an initial situation $S_0$, a natural language description of the starting state, and selecting which characters are present. Within each scene, writers generate \textit{beats} iteratively, with each beat capturing a unit of dramatic action. Applying beat $B_T$ to situation $S_{T-1}$ yields the updated situation $S_T$, forming a chain of causally connected story moments.

\subsection{Beat Generation}

\begin{figure}[!t]
  \centering
  \includegraphics[width=\columnwidth]{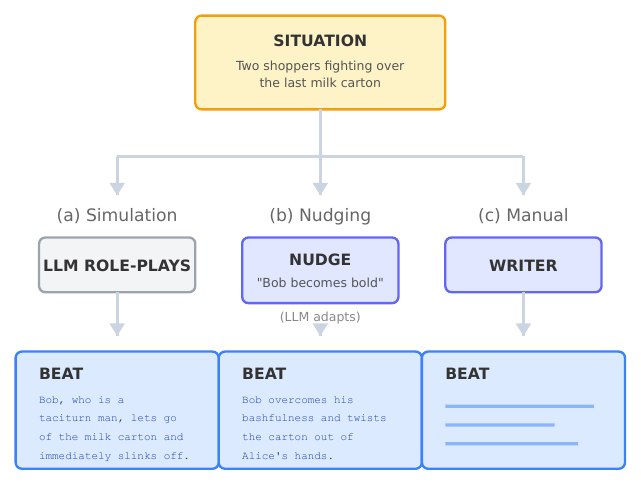}
  \Description{Diagram showing three beat generation mechanisms branching from a shared situation box reading Two shoppers fighting over the last milk carton. Branch a, Simulation: an LLM role-plays box leads to a beat card reading Bob, who is a taciturn man, lets go of the milk carton and immediately slinks off. Branch b, Nudging: a nudge box reading Bob becomes bold passes through an LLM adaptation step to a beat card reading Bob overcomes his bashfulness and twists the carton out of Alice's hands. Branch c, Manual: a writer box leads to a beat card with blank lines, indicating direct authoring.}
  \caption{\textnormal{Three mechanisms for beat generation. From a given situation, writers can:} \textbf{(a)~Simulation}\textnormal{: the LLM role-plays characters based on their traits and goals;} \textbf{(b)~Nudging}\textnormal{: the writer specifies a desired outcome, which the LLM adapts to; or} \textbf{(c)~Manual}\textnormal{: the writer authors beat content directly.}}
  \label{fig:beat-generation}
\end{figure}

TombWriter provides three mechanisms for generating beats (Figure~\ref{fig:beat-generation}), giving writers control over the balance between AI simulation and direct authorship:

\paragraph{Simulation.}
The LLM simulates the next beat by role-playing each character given their traits, goals, and the current situation. For example, given two shoppers fighting over the last milk carton: \texttt{Bob, who is a taciturn man, lets go of the milk carton and immediately slinks off.}

\paragraph{Nudging.}
The writer provides a free-text nudge specifying the desired outcome, and the LLM simulates how characters would arrive at that result. For example (nudge: ``Bob becomes uncharacteristically bold''): \texttt{Bob overcomes his bashfulness and twists the carton out of Alice's hands.}

\paragraph{Manual editing.}
The writer directly authors or edits beat content, with optional LLM assistance to polish the text. Any simulated beat can be manually revised to fine-tune the narrative.

Once the beat sequence is complete, prose can be generated from the story instrument based on the beat, allowing further manual editing, regeneration or changing intensity or style.

\subsection{Implementation}

TombWriter is implemented as a client-side web application using React 18 and TypeScript. State management uses Zustand with localStorage persistence, allowing writers to save and resume projects. For LLM integration, we use DeepSeek's API \texttt{deepseek-chat} (DeepSeek-V3.2) with user-adjustable temperature (0.1--2.0) to control generation variability. We selected DeepSeek-V3.2 for its strong performance on creative generation tasks and cost-effectiveness for a research prototype; the story archeology approach is model-agnostic. The system constructs prompts dynamically, incorporating character traits, memories, and narrative context.
\section{User Study}

We conducted a qualitative study to explore how experienced writers engage with beat-level story archeology when creating narratives with AI assistance.

\subsection{Study Design}

Initial pilot testing with novice writers using survey-based measures suggested that beat-level authoring supported creative engagement, but the quantitative data alone could not capture the nuanced ways writers experienced agency and ownership.
We therefore designed a qualitative study with experienced writers who could articulate their creative processes and compare TombWriter to their existing practices.

We recruited five participants from online creative writing and creative professional communities.
All had substantial experience: P01 and P04 were practiced writers working across multiple genres including fanfiction and original work; P02 was an experienced author actively working on a novel; P03 was a regular fiction writer; and P05 was a videographer and editor with storyboarding experience who writes original fiction.
All had prior experience with AI writing tools, including ChatGPT and Claude, as well as other AI-assisted creative writing tools.
Participants were compensated with gift cards.
We followed the ethics review requirements of our home institution.

\subsection{Procedure}

Participants attended an initial 30-minute onboarding session where we introduced the story archeology concept and demonstrated TombWriter's core features: character creation, beat generation (simulation, nudging, manual editing), and prose rendering.

Participants then used TombWriter independently over three days to develop a story of their choosing.
Unlike controlled lab studies, this extended engagement allowed writers to experience a larger arc of story development---returning to refine characters, discovering emergent plot directions, and iterating on beat sequences across multiple sessions.
We encouraged participants to work on stories they genuinely cared about rather than assigned prompts.

After the writing period, we conducted semi-structured interviews (45--60 minutes each) exploring:

\begin{itemize}
    \item Perceived control over narrative direction and character behavior  
    \item Sense of ownership over the resulting story
    \item Comparisons to their usual writing process and other AI tools
    \item Moments of surprise, friction, or creative discovery
\end{itemize}

Interviews were recorded, transcribed, and analyzed using thematic analysis~\cite{braun2006using}.
Two researchers independently coded initial transcripts, then collaboratively developed a codebook through discussion.
The final codes were organized into themes addressing our research focus on agency, ownership, and the experience of beat-level authoring.

\section{Findings}

Through semi-structured interviews following three days of independent TombWriter use, we identified four themes around agency, ownership, and the experience of beat-level authoring.

\subsection{Tool, Not Collaborator}

All participants framed AI as a generation engine requiring human direction, not a creative collaborator.
P04 articulated a clear criterion: \textit{``It would have to come up with its own ideas on an equal footing. It can't act without input, whereas I can write without the AI.''}
P02 offered a blunter framing: \textit{``Nobody says the hammer made the fence.''}
P05, who regularly uses AI tools for writing, emphasized the effort required: \textit{``You have to work at it a lot. You can't really just copy and paste—it's going to get it wrong.''}

This tool framing held even for participants who wanted more AI involvement.
P01, who preferred generating volume to select from, still described the relationship as human-driven: \textit{``It felt like my input was driving things.''}
The distinction was between wanting more output to curate versus wanting genuine initiative from the AI---the former remains compatible with tool framing.

However, participants valued the tool highly, reframing its purpose beyond writing assistance:

\begin{quote}
    \textit{``I don't see this as a writing tool, but rather as a creative discovery tool. Once you've developed an idea, it's very easy to lock yourself into a singular line of thinking. But there are infinitely many ways your story premise or character could develop. When I say `creative discovery tool,' I mean a tool that provides new perspectives you wouldn't have arrived at without external intervention.''} (P03)
\end{quote}

\subsection{Ownership and Voice}

All five participants claimed ownership of their output, but grounded it differently.
P02 applied a counterfactual test: \textit{``If not for me, would this exist? If the answer is no, then it is mine.''}
P04 tied ownership to vision adherence: \textit{``Getting the AI to stay close to that vision does help improve the sense of ownership.''}
P03 located ownership in the creative seed: \textit{``For me, the sense of ownership comes from whatever creativity I input directly.''}

Yet ownership was not uniformly strong.
P05 expressed hesitancy: \textit{``I kind of feel like I don't want to own it because it is so obvious. If I just copied and pasted this, I would expect to be called out.''}
This suggests ownership may depend on the perceived detectability of AI involvement: when the creative output reads as obviously AI-generated, claiming it becomes uncomfortable.

Multiple participants independently distinguished ownership from voice---a separation rarely made explicit in prior work.

\begin{quote}
    \textit{``Ownership is a separate concept from author voice. I felt like I was watching a tool write a story. It was technically mine, AI has no independent volition, but my voice was lost because the characters were not speaking or acting in ways that reflected how they behaved in my head.''} (P02)
\end{quote}

P03 concurred: \textit{``Your voice/style is lost almost entirely. But it's still yours, as it's your premise and they're your characters.''}
P05 experienced the voice problem acutely: \textit{``It's obviously AI... it feels like it's borrowing a lot from fan fiction.''}
This suggests ownership and voice are separable constructs.
Writers can feel a story is theirs while simultaneously feeling it does not sound like them, and when the voice gap becomes too large, even ownership becomes uncomfortable.

\subsection{Structure as Creative Value}

Participants consistently located TombWriter's value in structure and discovery rather than prose generation.
The system was described as \textit{``great for starting out''} and \textit{``faster than outlining by hand''} (P04), with P01 calling it a \textit{``starting machine''} for finding your way into a story.
P05, whose video editing background involves storyboarding, found the beat structure natural: \textit{``Just like you would storyboard for a movie, you storyboard for written narrative.''}

Several participants anticipated using TombWriter primarily for early-stage work.
P04 planned to use it \textit{``for the earlier stages, and swap to traditional once the story is nailed down.''}
P05 was explicit: \textit{``If I use this tool, I'm probably actually using it for a skeleton.''}
This aligns with P03's ``creative discovery tool'' framing: value lies in exploring narrative possibilities, not producing final prose.

The beat structure functioned as what P02 called \textit{``AI scaffolding.''}
Breaking the story into discrete chunks prevents the drift and inconsistency that plague longer AI generations.
As P04 noted: \textit{``With how often AI tends to go off the rails, it is better to keep them closer to what I described.''}
The structure also forced deeper engagement with character logic.
P05 found it \textit{``really forced me to sit and think about who my character is and how can I guide them to get the story that I want.''}

Participants adapted their use over the three days.
P04 learned to provide increasingly specific input after initial outputs diverged: \textit{``After it happened the first time I suppose I did get more specific with the beats.''}
P02 learned to \textit{``keep beats smaller''} to maintain consistency, describing it as resetting the context with each beat.

However, the structure also revealed a consistent boundary: control eroded at the prose layer.
P05 captured this precisely: \textit{``I felt really in control up until the last step. Until I actually saw the story, I was kind of like, `Eh.'''}
P01 found beat-level control too restrictive, preferring to \textit{``throw out a lot of it''} rather than micromanage.
The tradeoff between structural control and prose quality emerged as a key tension.

\subsection{Consistency and Surprise}

Participants wanted AI unpredictability, but bounded by story logic and character consistency.
P02 articulated the core principle: \textit{``The trick is that the surprises need to be consistent with the intention.''}

A key distinction emerged between execution-level and character-level surprise.
P02 explained: \textit{``If a sassy tsundere suddenly goes vulnerable and cries, it frustrates me. But if she does something mean and quips a hilarious insult I didn't expect, I'm there for it.''}
Unexpected choices in how a character acts were welcome; unexpected changes in who the character is were not.

P03 experienced positive surprise when the AI contributed context the writer had overlooked: \textit{``The religion aspect is one of the elements I totally forgot and the AI picked up for me.''}
It was an additive surprise that the AI filled gaps that fit the story's logic.

Surprise tolerance varied across participants.
P01 preferred high volume to select from; P04 preferred tight specification to minimize deviation.
But all shared the underlying principle: surprise should remain consistent with intention.

Multiple participants identified a tendency toward literalness and stereotyping.
P04 noted: \textit{``The AI focused a bit more on stereotyping the traits I gave for the characters, where I would play with those stereotypes but also subvert them.''}
P05 experienced similar typecasting: \textit{``It's going to typecast this character that you created.''}
The same mechanisms that prevent drift may also prevent the nuanced character development that experienced writers value.
\section{Discussion}

Our findings reveal how beat-level authoring shapes the experience of AI-assisted creative writing.
We discuss interaction patterns, the relationship between ownership and voice, and implications for designing human-AI writing interfaces.

\subsection{Interaction Patterns and Agency}

Participants developed distinct interaction strategies over the three-day study period, adapting their use of the system as they learned its capabilities and limitations.
Two patterns emerged: working at finer granularity (smaller beats, more specific descriptions) and strategic disengagement at the prose layer (treating output as skeleton rather than finished work).
Both patterns represent user agency expressed through interface adaptation where writers shaped their interaction with the system rather than accepting its defaults.

This adaptation resonates with Suchman's theory of situated action~\cite{Suchman1987}, which we invoked in framing beat-level authoring.
Expert writers used beats as resources for action rather than rigid scripts, adjusting granularity and specificity based on local conditions.
P04's shift toward more specific beats after initial failures, and P02's discovery that smaller beats ``reset the context,'' illustrate how writers developed practical strategies through use.

The consistent boundary at the prose layer---where participants reported control eroding---suggests a structural feature of beat-level authoring.
Writers maintain agency over narrative structure (what happens and why) while delegating surface realization (how it is phrased).
P05's observation that they felt ``really in control up until the last step'' captures this precisely.
For some writers, this division aligns with their creative priorities; for others, particularly those invested in prose craft, it represents a fundamental limitation.

\subsection{Ownership Without Voice}

A key finding is the distinction between ownership and voice, topics often combined in discussions of AI-assisted writing.
All five participants claimed ownership of their output, grounding it variously in counterfactual reasoning (P02), vision adherence (P04), creative seed (P03), or curatorial decisions (P01).
Yet multiple participants simultaneously reported that the output did not sound like them---their voice was lost even as their ownership was preserved.

This finding complicates prior work on AI writing and ownership.
Draxler et al.~\cite{draxler_ai_2024} identified an ``AI Ghostwriter Effect'' where writers do not perceive ownership of AI-assisted text and would not admit to AI involvement.
Our participants showed a different pattern: they claimed ownership but distinguished it from voice.
The ghostwriter framing assumes these constructs are unified; our findings suggest they can diverge.

Chen and Chan~\cite{chen2024large} found that ``ghostwriter mode'' (AI generating prose) caused anchoring effects and homogeneous outputs, while ``sounding board mode'' (AI providing feedback) better supported expert writers.
Beat-level authoring represents a third modality---structural scaffolding---where the AI neither writes prose directly nor comments on human prose, but instead simulates character behavior at an abstract level.
Our participants' willingness to claim ownership despite voice loss suggests this modality may navigate differently through the ownership landscape.

One possible mechanism for this preserved ownership is deliberate friction.
Unlike chat-based interfaces where a single prompt can generate hundreds of words, beat-level authoring requires writers to articulate characters, define scenes, and craft individual beats where each decision demanding explicit attention.
Dalsgaard~\cite{dalsgaard2025creative} argues that such friction in generative AI tools should not be optimized away but embraced: cognitive tension arising from the gap between intention and output can catalyze reflection, reframing, and deeper engagement.
In TombWriter, the friction is structural rather than accidental, as writers are required to specify before they generate, creating what Dalsgaard calls ``interpretive labor'' that binds them to the emerging work.
P02's observation that ``nobody says the hammer made the fence'' implicitly captures this: the fence belongs to its builder precisely because building required deliberate effort, not because the hammer performed optimally.

Yet friction alone may not be sufficient.
Professional playwrights in Grigis and De Angeli's longitudinal study~\cite{grigis2024playwriting} felt their creative effort was ``uniquely their own'' yet most ultimately rejected the AI technology.
Our findings suggest a possible mechanism: if voice is central to professional identity, ownership without voice may be insufficient.
P05's hesitancy to claim obviously AI-generated text points in this direction---when the voice gap becomes too visible, claiming ownership becomes problematic.

\subsection{Design Implications}

Our findings suggest several implications for designing AI writing interfaces.

\paragraph{Embrace tool framing, not collaborator framing.}
All five participants framed AI as a generation engine requiring human direction, not a creative collaborator.
This held regardless of whether they preferred tight specification (P04) or volume-based curation (P01).
The distinction matters for interface design: tools that position AI as a ``co-writer'' or ``creative partner'' may create expectations the technology cannot meet.
Our participants' clarity about the relationship---P02's ``nobody says the hammer made the fence''---suggests that transparent tool framing may better support productive use than aspirational collaborator framing.

\paragraph{Design for chunking as AI consistency management.}
Participants discovered that beat-level structure helps the AI maintain consistency, what P02 called ``AI scaffolding'' that ``resets the context'' with each beat.
This reframes chunking from a human organizational aid to an AI cognition management strategy.
The insight generalizes beyond writing: decomposing tasks into discrete, well-specified units may help maintain AI coherence across extended interactions.
Interfaces could make this mechanism explicit, helping users understand why granularity affects output quality.

\paragraph{Treat ownership and voice as separate design targets.}
Our participants claimed ownership while reporting voice loss, constructs often combined in AI writing discourse.
If ownership and voice can diverge, tools may need distinct mechanisms to address each.
Ownership in our study derived from structural decisions (characters, plot, direction); voice was lost at the prose layer.
Future tools might preserve ownership through structural control while offering additional mechanisms for voice---style controls, example passages, or iterative prose refinement.

\paragraph{Position AI writing tools for discovery, not production.}
Multiple participants described TombWriter as valuable for starting, outlining, and exploring narrative possibilities. A form of ``skeleton'' (P05) or ``starting machine'' (P01) rather than a prose production tool.
P03's reframe as a ``creative discovery tool'' captures this distinction.
Designing explicitly for early-stage creative work, with clear handoff points to traditional writing, may better match how experienced writers want to use AI assistance.

\paragraph{Preserve space for character complexity.}
The tension between consistency (preventing drift) and complexity (allowing character development) emerged across participants.
AI tended toward literalness and stereotyping---playing traits straight where human writers would subvert them.
Mechanisms that constrain AI behavior to prevent unwanted surprises may also prevent nuanced character development.
This tradeoff deserves explicit design attention, perhaps through user-controlled ``adherence'' settings or mechanisms for specifying intended character growth.

\subsection{Limitations and Future Work}

Our study has several limitations.
Five participants, while appropriate for qualitative exploration, cannot represent the full range of writer experiences.
All participants had prior AI tool experience, which may have shaped their expectations and strategies.
The three-day study period, while longer than typical lab studies, may not capture how practices evolve over extended use.

Our data on specific mechanism use (simulation versus nudging versus manual editing) was limited; participants did not systematically compare these options.
Future work should investigate how writers choose between interaction mechanisms and how these choices relate to ownership and agency.

The ownership-voice distinction emerged inductively from our data; future work should develop validated measures to study this distinction at scale.
Longitudinal studies tracking how ownership and voice perceptions evolve over extended collaboration would be particularly valuable.

Finally, our participants were experienced writers recruited from creative communities.
How novice writers experience beat-level authoring and whether the structural scaffolding provides different value for those still developing their craft remains an open question. Looking further ahead, beat-level authoring might extend to autonomous character agents with persistent goals and beliefs, incorporate role-playing mechanics for resolving dramatic conflicts, or adapt to other narrative forms including screenwriting and interactive fiction.

\begin{acks}
    This work was supported partly by Villum Investigator grant VL-54492 by Villum Fonden.
    Any opinions, findings, and conclusions expressed in this material are those of the authors and do not necessarily reflect the views of the funding agency.
\end{acks}

\bibliographystyle{ACM-Reference-Format}
\bibliography{tombwriter}

\end{document}